# Enhanced anomalous scattering by superconducting nanofilms vs T at O:K, Cu:$L_{2,3}$ , Ba:$M_{4,5}$ edges

J.V. Acrivos[(1)] SJSU, CA 95192-0101

*To Neville Francis Mott, mentor to solid state scholars on the 100th anniversary of his birth 9/21/1905*

## ABSTRACT

Enhanced (001) anomalous scattering by $YBa_2Cu_3O_{7-\delta}$ 50 nm films on $SrTiO_3$ substrates with and without a grain boundary versus temperature is interpreted using crystallographic weights to distinguish it from total electron yield and fluorescence spectra. The power of diffraction enhancement is to ascertain the film oxygen composition from the changes in the c-axis, $c_0$ as the film surface is scanned across the grain boundary, and to determine that $c_0$ is constant versus temperature across the superconducting phase transition.

(1) jacrivos@athens.sjsu.edu, TEL 408 924 4972, FAX 408 924 4945

## INTRODUCTION

Synchrotron X-ray absorption spectra (XAS) of layered cuprates, where superconducting planes are intercalated between ionic and perhaps magnetic layers in $YBa_2Cu_3O_7$ and derived phases ($YBCO_{x=6.5\ to\ 6.9}$) are compared at the O:K, Cu:$L_{2,3}$ and the Ba:$M_{4,5}$ edges. The film oxygen composition is obtained from the variation in the c-axis, $c_0$ that determines the (001) enhanced scattered amplitude.

## EXPERIMENTAL

The samples are 50 nm films, grown epitaxially by sputtering in an oxygen atmosphere onto a $SrTiO_3$ crystal with and without a 24 DEG ab grain boundary (GB) at the Complutense University and characterized by synchrotron XRD[1]. Spectra were collected versus photon energy, E at LBNL-ALS 6.3.1 station: by the (001) enhanced scattering ($I_s/I_0$) in the Kortright chamber at different temperatures[2] and distinguished from fluorescence ($F/I_0$) and total electron yield ($TEY/I_0$) in the Nachimuthu chamber where E was calibrated at E(CuO, Cu:$L_3$)=931.2eV[3]. A plane polarized beam (10 by 100 μm wide) of intensity $I_0$, incident on the 1cm$^2$ film at position x, at fixed angle θ to the film ab plane (FIG. 1) makes an angle 2θ with the detector, as reported in each spectrum (FIG. 2 - 4). The film samples are identified by whether the film is deposited on a single or a bi-crystal (SC or BC) qualified by the year fabricated/year measured. The oxygen composition is obtained by the comparison to XRD data[1,4,5].

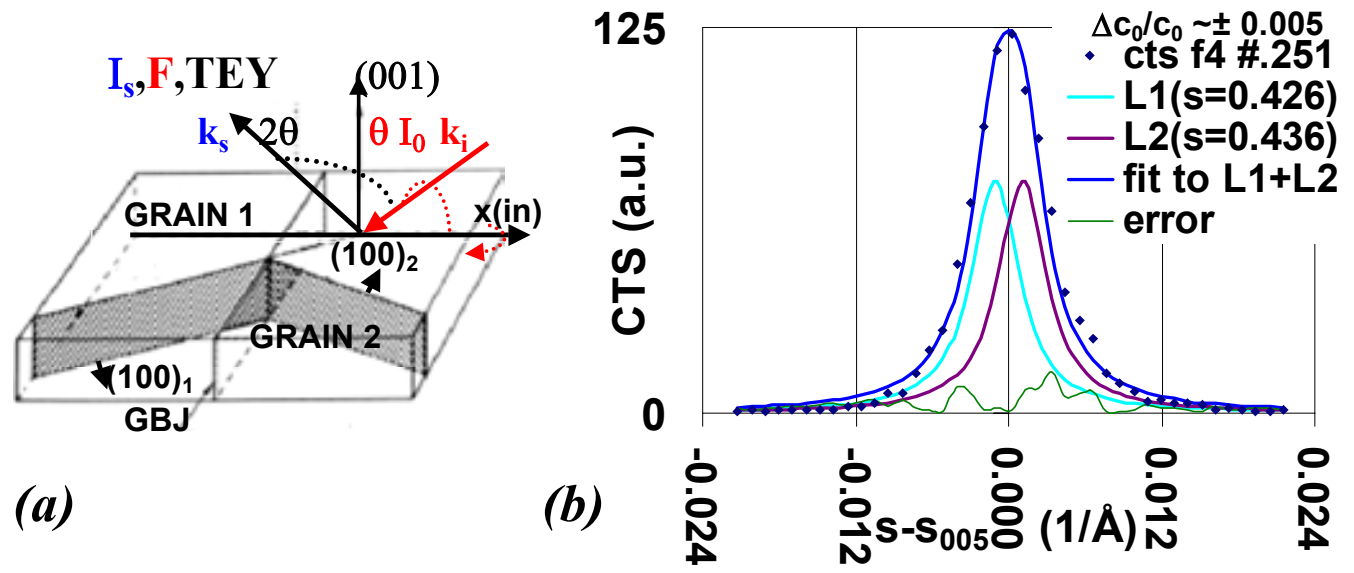

***FIG. 1:*** *Sample: **(a)** Measurement geometry determined by: the fixed horizontal incident beam $\mathbf{k_i}$, its position and angle θ by the sample displacement and rotation about the x-axis, and $\mathbf{k_s}$ by the detector angle 2θ to $\mathbf{k_i}$. **(b)** BC02/03 (001) XRD versus $s-s_{005} = 2\sin\theta/\lambda - 5/c_0$. The 100μm beam on GB detected two $c_0$ [ref. 1b].*

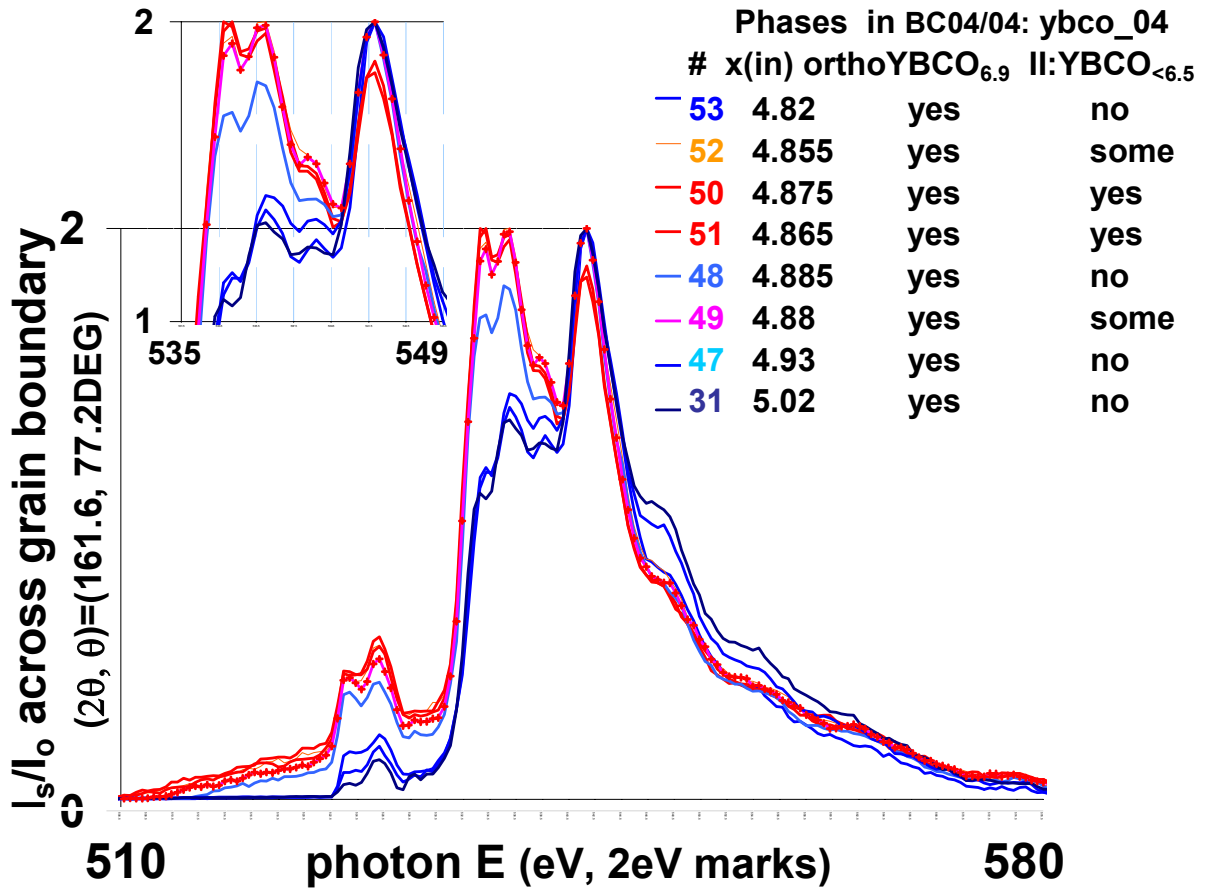

***FIG. 2:*** *Phases detected as incident beam position x goes through GB of film (FIG. 1a). Insert shows the appearance of two phases by the maximum change $\Delta E_{Bragg}/E_{Bragg} \approx -\Delta c_0/c_0$ from $E_{Bragg} \approx 546$ to $\approx 538$eV, or respectively [ref.5] $c_0 \approx 11.6$Å (ortho-I phase) increasing to $c_0 \approx 11.7$Å (ortho-II phase).*

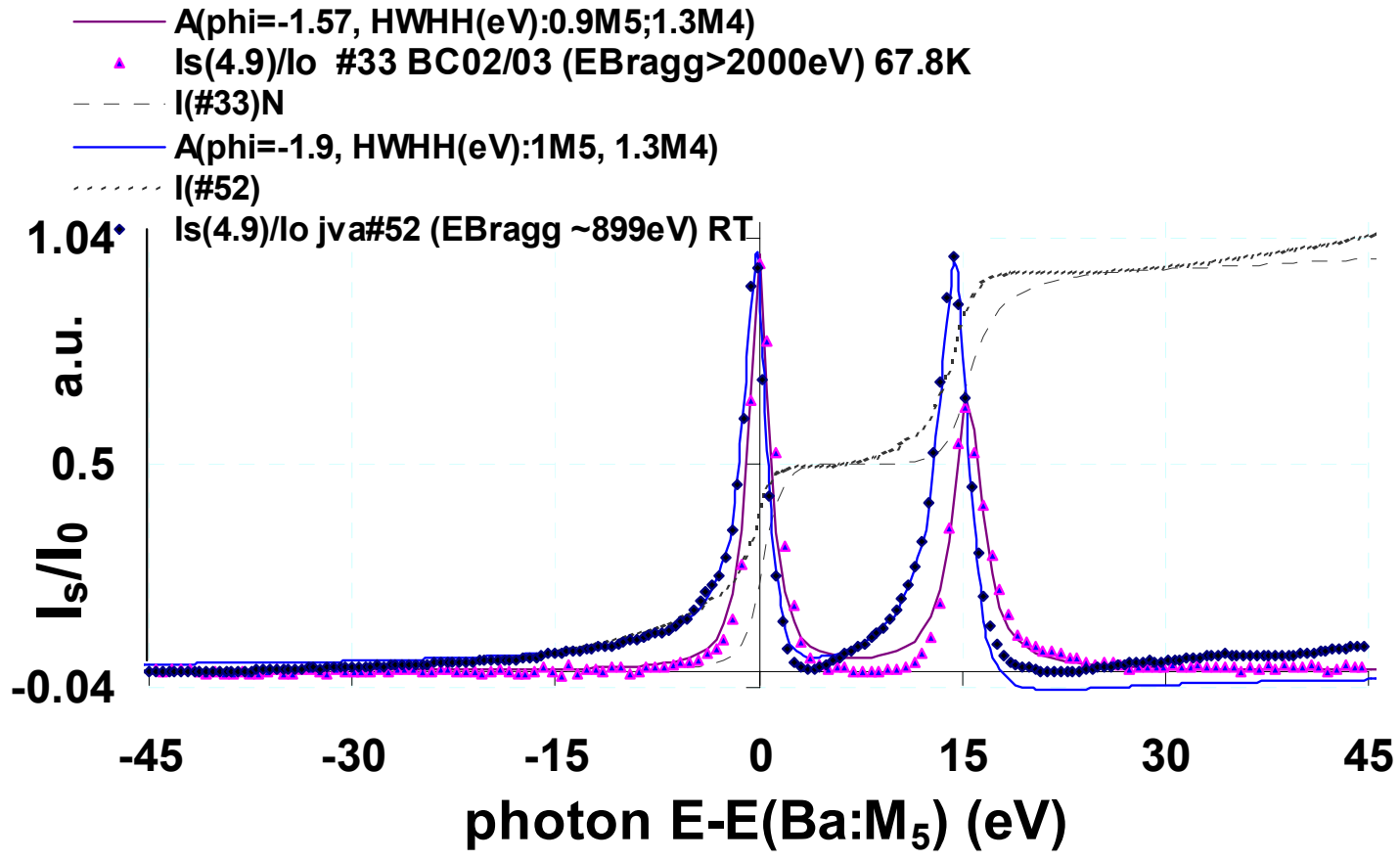

***FIG. 3:*** *Effect of $E_{Bragg}$ (2) on $I_s/I_0$ near the Ba:$M_{4,5}$ edges (BC02/03; BC04/04). Lifetime broadening and distortion due to some Ba, commonly occupying Y sites is observed. Broadening is evident in the integrated intensities, I from 730eV and the fit to A (4) indicates different HWHH at the $M_5$ and $M_4$ edge white lines, but the ratio of integrated intensities remain equal to one even as the lines narrow.*

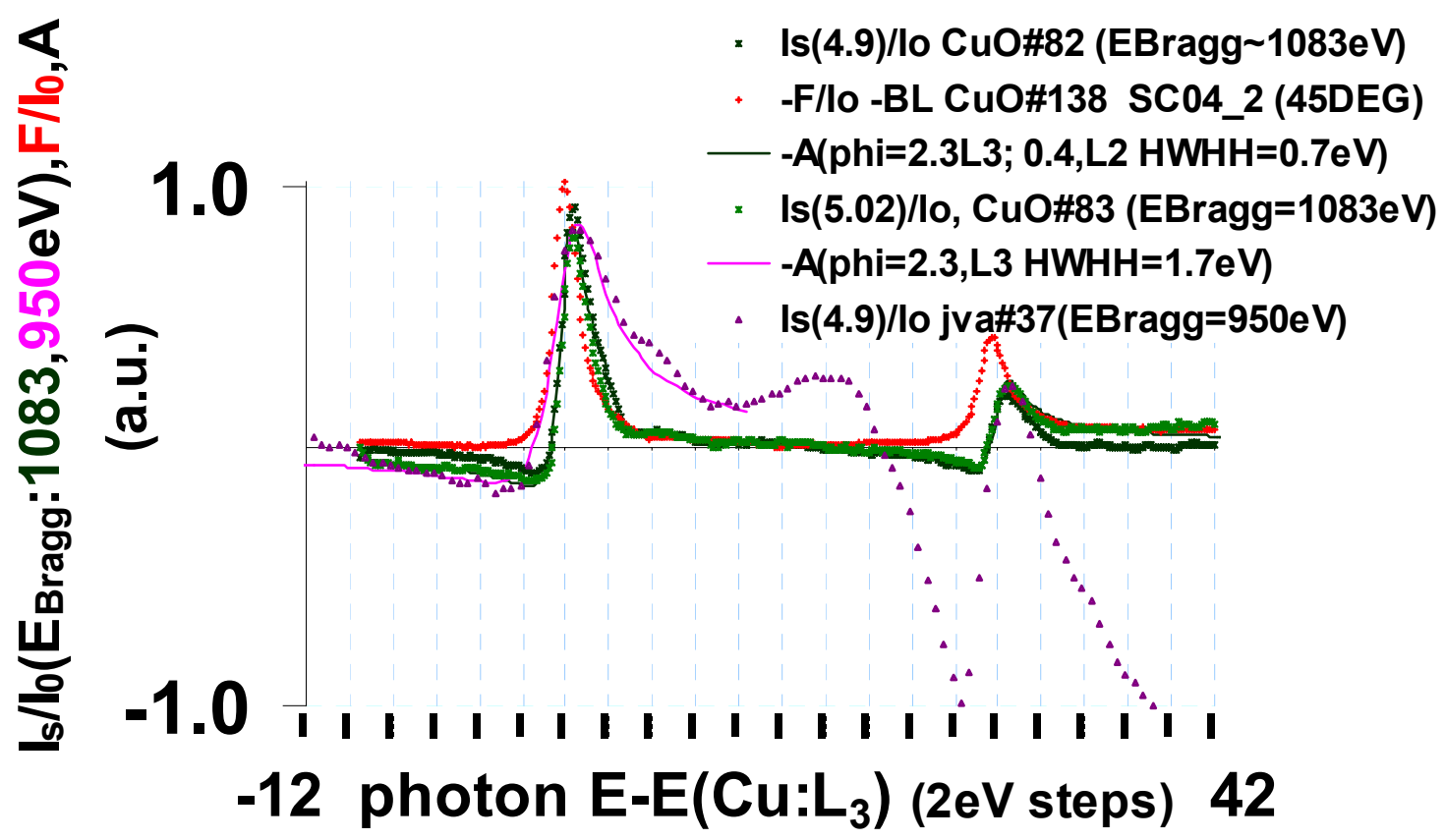

*FIG. 4: Effect of $E_{Bragg}$ on $I_s/I_0$ near the Cu:$L_{2,3}$ edges for film BC04/04 compared to $F/I_0$ and fitted to relation (4) $A(L_3)/A(L_2)=3$, $\alpha_{Cu}$ is constant to 1% in 40eV interval. 100eV broad background at $E(Cu:L_3)-E_{Bragg}$ ~18eV disappears at $10^2$eV.*

## DISCUSSION

In the soft X-ray region only the enhancement of the (001) diffraction is accessible. The scattering amplitude by atoms j depends on the incident and scattered photon: momenta $\mathbf{k_i}, \mathbf{k_s}$ (FIG. 1a), polarization $\hat{e}_i, \hat{e}_s$ and energy E [4, 6]:

$$f_j(\mathbf{k_i}, \mathbf{k_s}, E) = f_j^0(\mathbf{k_i}, \mathbf{k_s}) + \Delta f_j(\mathbf{k_i}, \mathbf{k_s}, \hat{e}_i, \hat{e}_s, E). \quad (1)$$

The Thomson amplitude $f_j^0$ is the matrix element of the square of the vector potential acting on the electron number density. The anomalous amplitude near an absorption:

$$\Delta f_j(\mathbf{k_i}, \mathbf{k_s}, \hat{e}_i, \hat{e}_s, E) = f_j' + i\, f_j'' \approx$$

$$\Sigma_j \Sigma_{nl}[<\hat{\mathbf{e}}_s.\mu_{ln}\, e^{-i\mathbf{k}_s.\mathbf{r}_j}><e^{i\mathbf{k}_i.\mathbf{r}_j}\, \mu_{nl}^*.\hat{\mathbf{e}}_i>]/[E_n-E_l+E+(\Delta_n - i\ \mathrm{HWHH})] + \mathrm{HC}$$

involves dipole matrix elements $\mu_{ln}$ between initial and final states (n,l) with energies $E_n$, $E_l$, that depend on: orientation in a layer cuprates[7] where the incident $\varepsilon_{X\text{-ray}}$ unit vector, $\hat{\mathbf{e}}_i$, is in the film ab plane, the half width at half height HWHH determined by lifetime broadening, the crystallographic site diffraction weights $\alpha_j = \Sigma_j e^{i(\mathbf{k_i}-\mathbf{k_s}).\mathbf{r_j}}$ where $(\mathbf{k_i}-\mathbf{k_s}).\mathbf{c_0} = 2\pi\, E/E_{Bragg}$ for (001) diffraction, $\Delta_n$= Lamb shift, $f_j'$= dispersion, $f_j''$= absorption and HC=Hermitean conjugate[6]. The $YBCO_x$ (001) diffraction enhancement depends on α and E through the Bragg relation:

$$E_{Bragg} = hc/2\sin(\theta)/c_0 \ + \text{Stenström correction}, \quad (2)$$

determined by the magnitude of the c-axis, $c_0$, h= Planck constant, c= velocity of light. Hanzen[5a] has correlated the oxygen composition of $YBCO_{x=7-\delta}$ with $c_0$ in each phase. Thus, at fixed orientation minute changes:

$$\Delta E_{Bragg}/E_{Bragg} \sim -\Delta c_0/c_0. \quad (3)$$

detect the appearance of new phases as the incident beam scans the film surface with fixed $\mathbf{k_i}$, $\mathbf{k_s}$ (FIG. 1a).

As E=>$E_{Bragg}$ the edge white line, WL lifetime broadening increases and/or enhanced Compton and Rayleigh scattering is observed. As $E_{Bragg}$-E increases, the tail of $f^0$ becomes a baseline correction (FIG. 2 - 4), but the sample may rotate the plane polarized light by an angle ϕ. The signal and its Hilbert-Kramers-Kronig transform[4]:

$$\begin{pmatrix} I_s/I_0(E) \\ P\int_{-\infty}^{\infty} \frac{I_s/I_0(E')}{\pi(E'-E)}\, dE' \end{pmatrix} = \begin{pmatrix} \mathrm{Cos}(\phi) & -\mathrm{Sin}(\phi) \\ \mathrm{Sin}(\phi) & \mathrm{Cos}(\phi) \end{pmatrix} \begin{pmatrix} f' \\ f'' \end{pmatrix}$$

are a mixture of real and imaginary terms in (1)[4, 6]. The data (FIG. 2 - 4) are analyzed with the purpose to ascertain the film properties for important useful as follows:

**(i)** A mixture of real and imaginary components is observed in the WL at the Cu:$L_{2,3}$ and Ba:$M_{5,4}$ (FIG. 3-5) for BC and SC films. The TEY/$I_0$ and F/$I_0$ show a Lorentz type shape WL with an edge jump weaker than 1% of its amplitude[7]. Thus if the $f^0$ tail is linear, the enhanced scattered amplitude minus a base line may be compared to:

$$A_j = I_{sj}/I_0/\alpha_j = [y\,\mathrm{Cos}(\phi) - \mathrm{Sin}(\phi)]/[1 + y^2] \quad (4)$$

where $y=(E-E_0)/\mathrm{HWHH}$, $E_0$ is the edge energy and HWHH is the WL half width at half height. The data are normalized to a constant α, and fitted to the minimum and maximum amplitude in (4) to over determine ϕ ($A_{max}=\mathrm{Sin}^2(\phi/2)$, $A_{min}=-\mathrm{Cos}^2(\phi/2)$, $y(A_{max})-y(A_{min})=2/\mathrm{Cos}(\phi)$; $y(A=0)=\mathrm{Tan}(\phi)$). The fitted A (FIG. 3-5) indicate that the sample rotates the plane polarized beam by $\phi(Cu:L_3)\approx 3\pi/4\pm\pi$ at $E(Cu:L_3)-E_{Bragg}$= 18 to $10^2$eV. $I_s/I_0[E(Cu:L_2)\approx E_{Bragg}]\approx f'$ (FIG. 4) agrees with theory $\phi \approx 0$[4,6]. Lifetime broadening[6c] is observed as $E_{Bragg}$-$E(Cu:L_3)$ decreases from $10^2$ to 18eV, with respective HWHH=0.7 to 1.7eV. ϕ can not be determined from the fit to (4) alone. The WL (FIG 4, 5) absorption at the $L_{2,3}$ edges depends on the Cu valence:

$Cu^{1+}(.2p^6..) \Leftrightarrow Cu^{+1}(.2p_{3/2}{}^3..4d_{5/2})$ ; $Cu^{+1}(.2p_{1/2}..4d_{3/2})$,
$Cu^{2+}(.\,2p^6 3d^9) \Leftrightarrow Cu^{+2}(.2p_{3/2}{}^3.3d^{10})$; $Cu^{+2}(.2p_{1/2}..4d_{3/2})$,
$Cu^{+3}(.2p^6 3d^8) \Leftrightarrow Cu^{+3}(.2p_{3/2}{}^3..3d^9)$; $Cu^{+3}(.2p_{1/2}..3d^9)$,

the crystal field splitting (different at the $L_2$ and $L_3$ edges) and orientation[7-9] making it difficult to assign spectral features to the Cu sites in $YBCO_x$ where the local crystal field symmetry near the Cu:2 site is square planar in the ab-plane while near the site Cu:1 it is square planar in the bc-plane. The advantage of scans at different fixed orientations is that site identification is made by changes in $\alpha_j(E) = n_j\cos(2\pi\, z_j E/E_{Bragg})$ for different fixed $E_{Bragg}$, when $n_j$ is the number of equivalent atom j sites with coordinate $z_j$ in the unit cell (Table I). When $E(Cu:L_2)=E_{Bragg}\approx$ 950eV, $\alpha_{Cu:2}/\alpha_{Cu:1}\approx$ -1.4 the enhanced shoulders ~ 6eV above the main signal, but of opposite sign amplitude may be due to the Cu:1 site contribution. Since the exact cancellation expected for $E_{Bragg}$= 1083eV, $\alpha_{Cu:1}/\alpha_{Cu:2}\approx$-1 (if the second order matrix elements in $\Delta f_{Cu}$ for both sites are of the same order of magnitude) is not observed, the contributions from Cu:1 and Cu:2 sites are different and appear at different E, which indicates the sites are occupied by different valence copper and $\phi\approx 3\pi/4$ in (4).

**(ii)** The data at the O:K edge (FIG. 2) indicates that a displacement of the 100µm wide beam, across the GB detects a new enhancement peak, associated with a higher $c_0$ phase. Its relative intensity (FIG. 2) in the extended X-ray absorption region (XAFS) centered at 538eV ($c_0\approx$11.7Å) identifies it with the ortho-II phase ($YBCO_{<6.5}$) relative to that at 546eV ($c_0\approx$11.6Å) for the ortho-I phase ($YBCO_{\approx 6.9}$) in agreement with the XRD data[1b] (FIG. 1b). The width of the GB is estimated to be comparable to the beam width since the enhancement, at $E_{Bragg}\approx$538eV appears only within x≈ 4.87±0.005in, while that at 546eV changes very little across the GB. The relative intensities can not be measured due to strong enhancement dependent $I_s/I_0$ near $E_{Bragg}$. $YBCO_x$ $c_0$ data versus oxygen composition indicates that near the GB a ~5% O decrease induces the ortho-II phase where $k_x$=-$k_y$ periodic lattice distortions (PLD) release the film strain[1c,d,5b,8c].

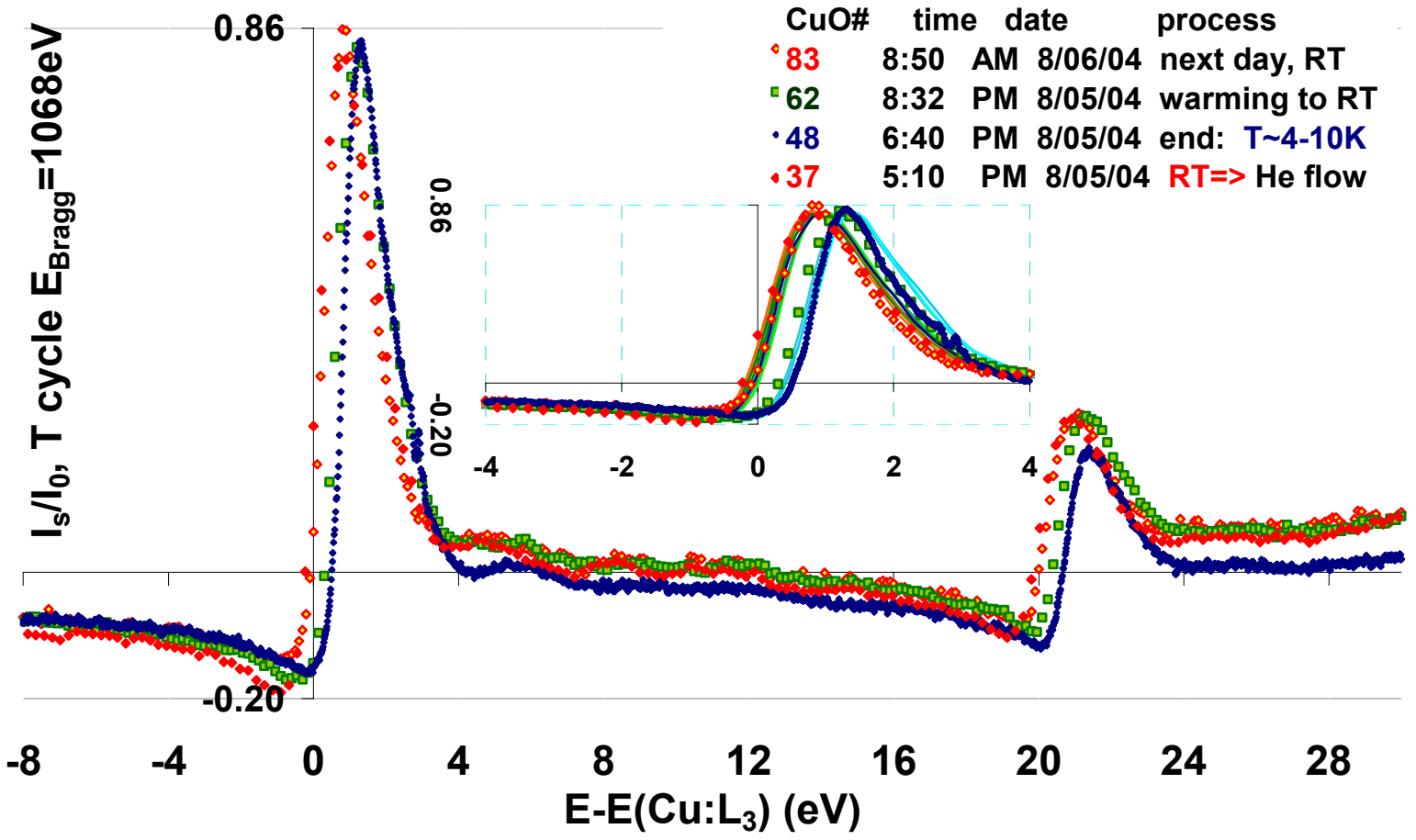

*FIG. 5: Cu:$L_{2,3}$ edge $I_s/I_0$ (BC04/04, x=5.02in) at different temperatures, T and fixed $E_{Bragg}$=1083eV. Insert shows the superposition of scans 37 to 62 as T decreases across $T_c$. Two temperature intervals are identified by scans taken 5min. apart as T goes through $T_c$, showing first a reversible shift of 0.5eV to higher energy, then narrowing from HWHH=0.7 to 0.5eV.*

**Table I**: Assignment of the $YBCO_x$ unit cell sites (Hanzen notation[5a]) by correlation of $I_s/I_0$ to crystallographic diffraction weights α when $E_=$=$E_{Bragg}$, $E_{\neq}$≠$E_{Bragg}$ (FIG. 2- 6).

| $YBCO_x$ | $D^{17}_{2h}$ | Site: | O:1 | O:2 | O:3A,B | Cu:1 | Cu:2 | Ba | Y |
|---|---|---|---|---|---|---|---|---|---|
| (001) enhancement | | z: | 0 | ± .18 | ± .378 | 0 | ± .358 | ± .18 | ± .5 |
| | | α($E_=$): | 1 | 0.8 | -2.9 | 1 | -1.4 | 0.8 | -1 |
| $I_s/I_0$ Peak | | α($E_{\neq}$): | 1 | 1.9 | 3.4 | 1 | −1 | 1.9 | 0.6 |
| E (eV) | Sign $I_s/I_0$ | | Site Contribution to Signal | | | | | | |
| 528 | + | + | | yes | | | | | |
| 530 | + | | yes | | | | | | |
| 531-535 | - | | | | yes | | | | |
| 531-535 | | + | | | yes | | | | |
| 538 | + | + | maybe | yes | | | | | |
| 545 | + | + | | | yes | | | | |
| 779 | + | + | | | | | | yes some Ba | |
| 793 | + | + | | | | | | yes some Ba | |
| 932 | ± | ± | | | | | yes | | |
| 950 | -± | ± | | | | | yes | | |
| 936 | -± | | | | | yes | | | |
| 954 | -± | | | | | yes | | | |

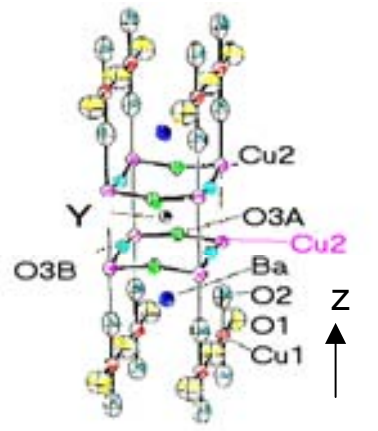

The effect of buckling is observed in single crystal films (FIG. 6c #58; 56). Satellites appear next to the 545eV enhanced peak in SC04/04 $YBCO_{6.9}$, in the XAFS region at $E_{Bragg}$= 542, 551eV where the tilt in θ follows from the Bragg relation (2), $\Delta\theta/\tan(\theta)$ = $-\Delta E/E$ = ± 0.008 or $\Delta\theta \approx \pm$ 0.04. Evidence of lifetime broadening in $I_s/I_0$ is obtained by comparison with TEY/$I_0$ and F/$I_0$ in similar orientation (FIG. 6c), E=528, 530eV for BC and SC films.

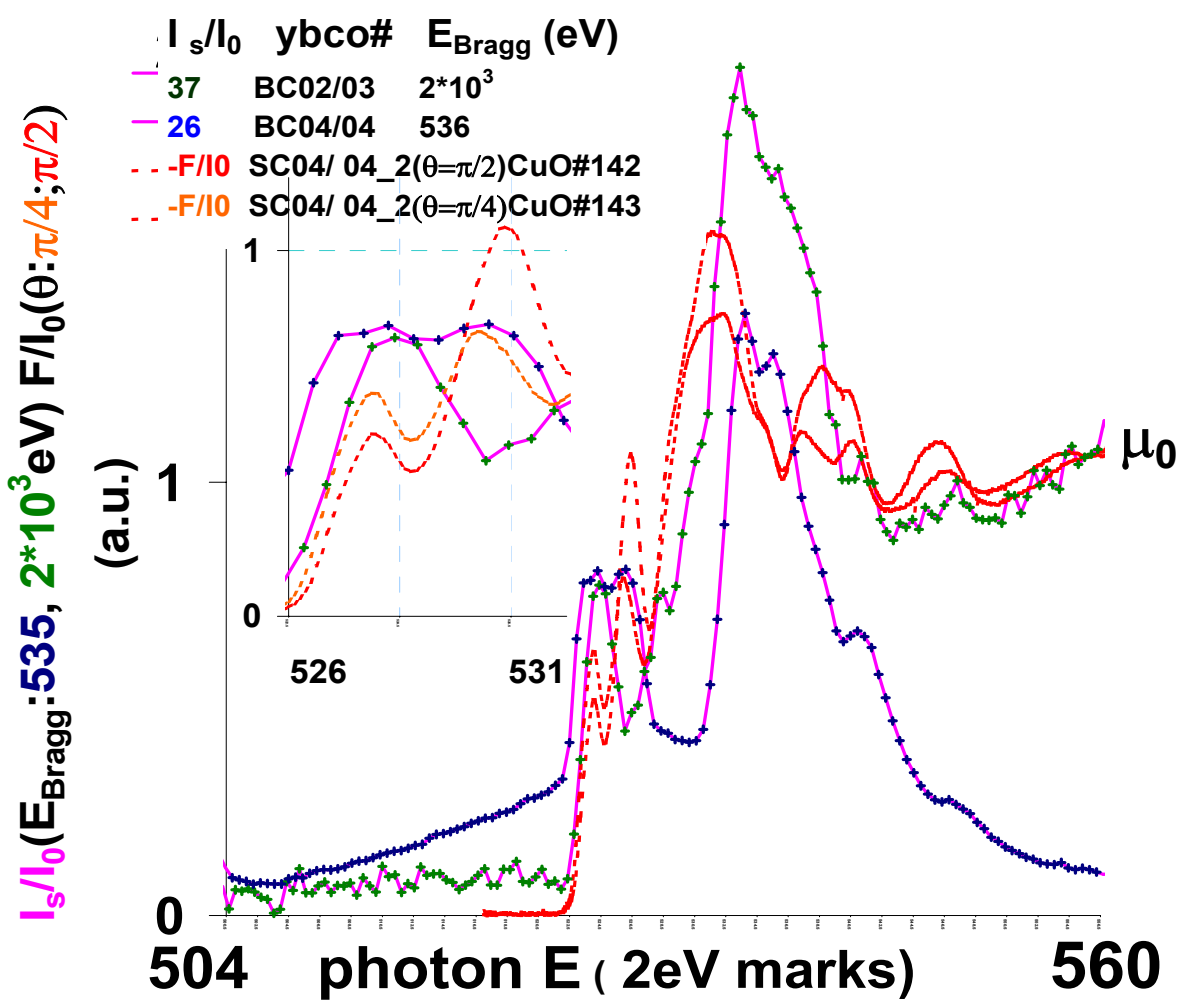

*FIG. 6: O:K edge comparisons: **(a)** $I_s/I_0$($E_{Bragg}$≈ 536; 2*$10^3$eV) and F/$I_0$(θ = π/4, π/2).*

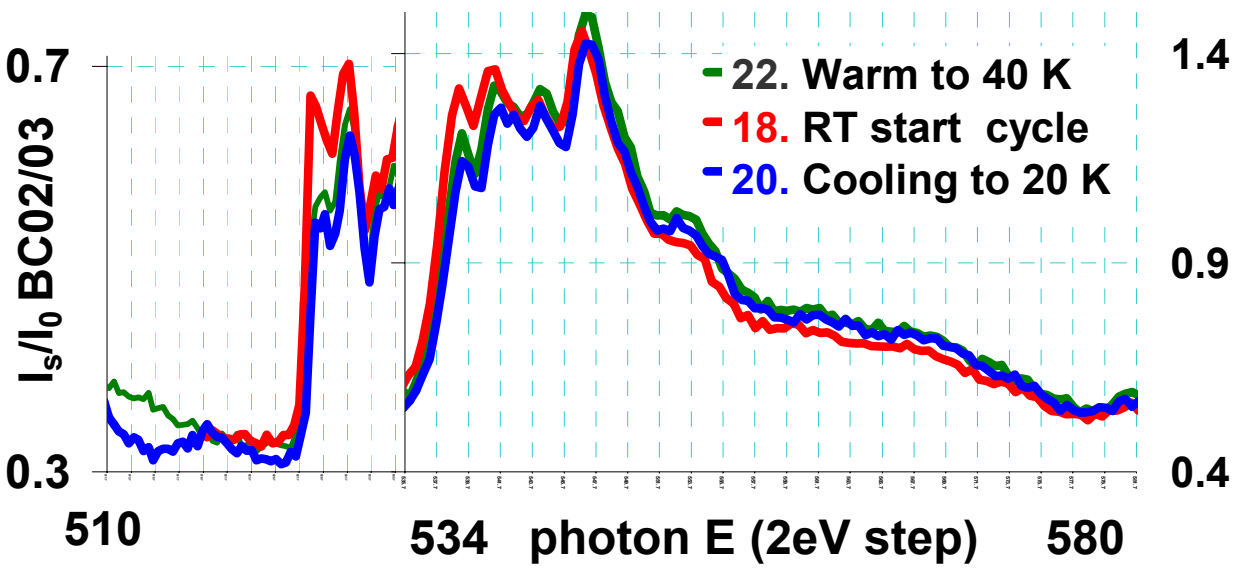

***(b)** $I_s/I_0$ through $T_c$ cooling cycle. Only the broad background centered near 528eV decreases across $T_c$.*

The new information obtained by the comparisons (FIG. 6) in similar orientation is on the K-edge transitions:

$$O{:}1s^2 \Leftrightarrow O{:}1s.np_{x,y},\ n>2 \text{ and } O{:}1s^2 \Leftrightarrow O{:}1s.np_z,\ n>2.$$

The first are allowed when $\hat{e}_i$ is in the ab plane and the second when it is normal according to relation (1). The (001) $I_s/I_0$ enhancement near 528eV is identified with the O:2 site (Hanzen notation[5a]) because it increases with α(O:2) when $E_{Bragg}$ increases from 545 to 2*$10^3$eV. The amplitude maximum in F/$I_0$ and TEY/$I_0$ remains independent of orientation as is expected in a nearly local octahedral field (FIG. 6). The enhancement near 530eV is identified with site O:1 because it looses relative intensity as α(O:2)/α(O:1) goes from 0.8 to 1.9 for the different $E_{Bragg}$, above with $\hat{e}_i$ in the ab-plane. $I_s/I_0$(E=531 to 536eV) and different $E_{Bragg}$ (FIG. 6) is identified with site O:$3_{A,B}$ absorbance because it changes sign with α(O:$3_{A,B}$):

$$I_s/I_0(E_{Bragg} \sim 546eV,\ \alpha(O{:}3_{A,B}) \approx -2.7) \leq 0$$

$$I_s/I_0(E_{Bragg} \sim 2{*}10^3eV,\ \alpha(O{:}3_{A,B}) \approx 3.5) > 0$$

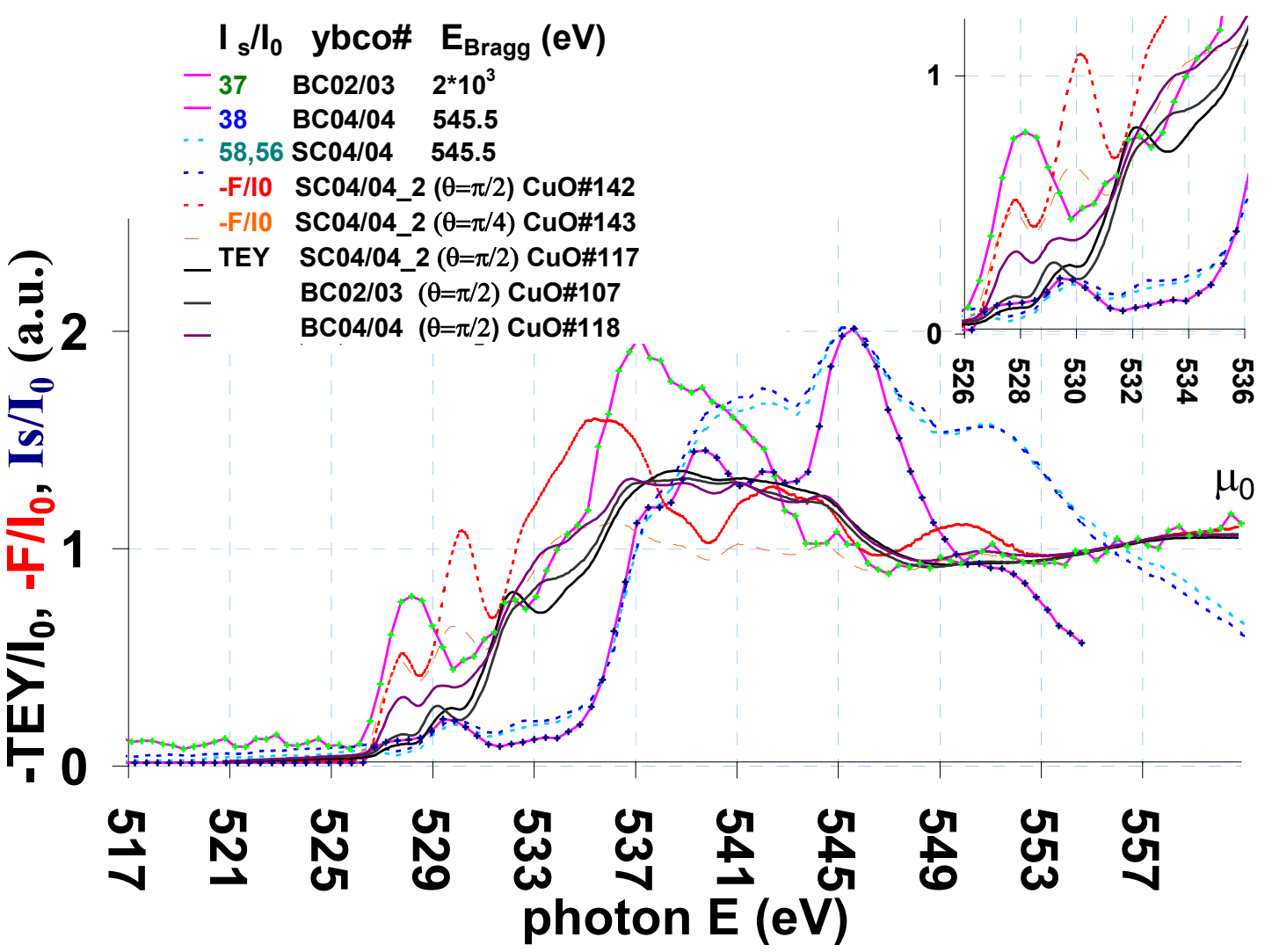

***6(c)*** *$I_s/I_0$, $F/I_0(\theta=\pi/2)$ for BC and SC films. TEY/$I_0$ is constant across GB for BC04/04 except BC02/04 that shows loss of O.*

The assignments made by a single diffraction in the soft X-ray region using the variation of α are similar to those made by different measurements[8] for de-twinned single crystals $YBCO_x$, and may be correlated to the chemical valence: The most negative ionic valence is associated with the lowest edge energy for site O:2 in the BaO layer, the next higher edge energy is for site O:1 in the CuO chains and finally the highest valence is assigned to sites $O{:}3_{A,B}$ where molecular orbital calculations show that the $CuO_2$ layer in $YBCO_x$ nano-particles is covalent with a Mulliken atomic charge at the $O{:}3_{A,B}$ sites of –1.3 and 0.8 at the Cu:2 sites[1c, 9].

The temperature dependence obtains a constant $c_0$ by the unchanged enhancement peak at $E_{Bragg}(O{:}K) \approx 546$eV for the ortho-I phase (FIG. 6b). $\phi(Cu{:}L_{3,2}) = 3\pi/4$ in (4) (FIG. 5) is constant across $T_c$ but the observed edge shift by 0.5eV at the $Cu{:}L_{3,2}$ edges is assumed to be due to increased ionization at the Cu:2 sites in $CuO_2$ plane below $T_c$.

Enhancement at $E_{Bragg}$= E(O:K edge XAFS) may be due to resonance shake-up absorption up to 20eV above the edge[7-9] identified by $\alpha(E=E_{Bragg})$ and $\alpha(E\neq E_{Bragg})$ (Table I, FIG. 2, 6). Enhancement at 536eV is associated with O:1 and/or O:2 sites by the increase in α. Enhancement at 545eV is associated with $O{:}3_{AB}$ through the negative XAFS amplitude $F/I_0-\mu_0$. This suggests that the first are associated with the ortho-II phase and the latter with the ortho-I, in agreement with an oxygen concentration decrease in O:1 sites in ortho-II phase and PLD formation[5b,8c]. Anisotropy is evident both through the sample properties and the crystallographic weights.

**(iii)** In the ideal structure, Ba occupies a unique site in $YBCO_x$ (Table I) but in real crystals it also occupies Y sites. As $E(Ba{:}M_{4,5})$=> $E_{Bragg}$ the (001) $I_s/I_0$ WL enhancement shows lifetime broadening with $\alpha_{Ba}/\alpha_Y$=-0.8 but for $E_{Bragg}$-$E(Ba{:}M_{4,5})> 2*10^3$eV, $\alpha_{Ba}/\alpha_Y$=>3 and the WL narrows (FIG. 3 #52 and #33). The expected ratio of amplitudes for transitions:

$$Ba^{2+}(.3d^{10}.) \Leftrightarrow Ba^{+2}(.3d_{5/2}{}^{5}.4f_{7/2})\ ;\ Ba^{+2}(.3d_{3/2}{}^{3}.4f_{5/2}),$$

proportional to initial state multiplicities, $A(M_5)/A(M_4)$= 1.5 is approached only for $E_{Bragg} > 2*10^3$eV. A comparison of TEY/$I_0$ for BC04/04 with a standard $BaBr_2$ shows that the relative intensities for the WL at the two edges are orientation dependent[7c, 9].

In summary, although only the (001) diffraction enhancement $I_s/I_0$ is accessible in the soft X-ray region, the crystallographic weights α at different $E_{Bragg}$ also provide new valuable information, as in the hard X-ray region[6].

## CONCLUSION

The (001) anomalous enhanced scattering in $YBCO_x$ nano-films is very sensitive to the magnitude of the film c-axis and consequently the O composition in fabricated nano-films and should therefore be used to explain the properties of each in order to characterize their properties for important industrial use.

## ACKNOWLEDGEMEMNTS

Work was supported by the NSF and Dreyfus Foundations at SJSU, and DOE at LBNL-ALS and all colleagues[1-9]. All solid state scholars are grateful to the memory of N. F. Mott on his 100th birthday.